\documentclass[]{aa}

\usepackage{amsmath}
\usepackage{amssymb}
\usepackage{url}
\usepackage{graphicx}
\usepackage{times}
\usepackage[utf8]{inputenc}
\usepackage{xspace}
\usepackage{color}
\usepackage{multirow}
\usepackage{multicol}
\usepackage{float}

\bibpunct{(}{)}{;}{a}{}{,}

\usepackage{etoolbox}

\title{Accretion-to-jet energy conversion efficiency in GW170817}
\titlerunning{Accretion-to-jet energy conversion efficiency in GW170817}
\author{O.~S.~Salafia\inst{\ref{oab.me},\ref{infn.mib}} and B.~Giacomazzo\inst{\ref{unimib},\ref{infn.mib},\ref{oab.me}}}

\institute{INAF -- Osservatorio Astronomico di Brera, via E. Bianchi 46, I-23807 Merate (LC), Italy\label{oab.me} \and INFN -- Sezione di Milano-Bicocca, Piazza della Scienza 3, I-20126 Milano (MI), Italy\label{infn.mib}
\and Università degli Studi di Milano-Bicocca, Dip. di Fisica ``G. Occhialini'', Piazza della Scienza 3, I-20126 Milano, Italy\label{unimib}}

\authorrunning{Salafia \& Giacomazzo}

\date{Received 5 May 2020 / Accepted 21 October 2020 / Corrected 25 March 2022 }

\abstract{Gamma-ray bursts (GRBs) are thought to be produced by short-lived, supercritical accretion onto a newborn compact object. Some process is believed to tap energy from the compact object, or the accretion disc, powering the launch of a relativistic jet. For the first time, we can construct independent estimates of the GRB jet energy and of the mass in the accretion disc in its central engine; this is thanks to gravitational wave observations of the GW170817 binary neutron star merger by the Laser Interferometer Gravitational wave Observatory (LIGO) and Virgo interferometers, as well as a global effort to monitor the afterglow of the associated short gamma-ray burst GRB 170817A on a long-term, high-cadence, multi-wavelength basis. In this work, we estimate the accretion-to-jet energy conversion efficiency in GW170817, that is, the ratio of the jet total energy to the accretion disc rest mass energy, and we compare this quantity with theoretical expectations from the Blandford-Znajek and neutrino-antineutrino annihilation ($\nu\bar\nu$) jet-launching mechanisms in binary neutron star mergers. Based on previously published multi-wavelength modelling of the GRB 170817A jet afterglow, we construct the posterior probability density distribution of the total energy in the bipolar jets launched by the GW170817 merger remnant. By applying a new numerical-relativity-informed fitting formula for the accretion disc mass, we construct the posterior probability density distribution of the GW170817 remnant disc mass. Combining the two, we estimate the accretion-to-jet energy conversion efficiency in this system, carefully accounting for uncertainties. The accretion-to-jet energy conversion efficiency in GW170817 is $\eta\sim 10^{-3}$, with an uncertainty of slightly less than two orders of magnitude. This low efficiency is in agreement with expectations from the $\nu\bar\nu$ mechanism, which therefore cannot be excluded by this measurement alone. 
The low efficiency also agrees with that anticipated for the Blandford-Znajek mechanism, provided that the magnetic field in the disc right after the merger is predominantly toroidal (which is expected as a result of the merger dynamics). This is the first estimate of the accretion-to-jet energy conversion efficiency in a GRB that combines independent estimates of the jet energy and accretion disc mass.
Future applications of this method to a larger number of systems will reduce the uncertainties in the efficiency and reveal whether or not it is universal. This, in turn, will provide new insights into the jet-launching conditions in neutron star mergers.} 

\keywords{relativistic processes, gamma-ray burst:individual -- GRB170817A, stars:neutron, gravitational waves}

\begin{document}

\maketitle

\section{Introduction}

It has been long established that the outflows that produce gamma-ray bursts (GRBs) must expand at relativistic speeds \citep[e.g.][]{Ruderman1975,Goodman1986,Paczynski1986,Fenimore1993,Woods1994,Woods1995,Frail1997,Goodman1997,Taylor2004} and be collimated \citep[e.g.][]{Rhoads1997,Meszaros1999,Sari1999,Ghisellini1999,Rhoads1999,Frail2001}, that is, they must be relativistic jets. The widely accepted site for the production of such jets is an accreting stellar-mass compact object \citep{Piran2004}, either a black hole (BH) or a neutron star \citep{Usov1994,Bucciantini2008}. The first association of a supernova with a long GRB \citep{Galama1998} strongly supported this link, pointing to the gravitational collapse of a massive star as the progenitor. More recently, the association \citep{LVC2017GRB,LVC2017multimessenger} of the gravitational wave event GW170817 \citep{LVC2017discovery}, interpreted as having been produced by the merger of a binary of neutron stars (BNS hereafter, \citealt{LVC2017discovery,Abbott2019}) with the short gamma-ray burst (SGRB) GRB 170817A \citep{Goldstein2017,Savchenko2017}, confirmed the long-held expectation \citep{Bisnovatyi-Kogan1975,Paczynski1986,Eichler1989} that some GRBs are produced in compact binary mergers. The identification of an optical counterpart to GW170817 \citep{Coulter2017,Valenti2017}, later spectroscopically classified \citep{Pian2017,Smartt2017} as a kilonova \citep{Li1998,Metzger2016}, pinpointed the host galaxy of the event, allowing for a long-term, multi-wavelength monitoring of its location. This uncovered an additional non-thermal counterpart \citep[e.g.][]{Hallinan2017,Margutti2017,Troja2017,Alexander2017,Alexander2018,DAvanzo2018,Margutti2018,Dobie2018,Lyman2018,Lamb2019,Hajela2019} that was eventually established \citep[][thanks to very-long-baseline interferometry imaging]{Mooley2018a,Mooley2018b,Ghirlanda2019} as being the afterglow (i.e.~synchrotron emission from the external shock in the interstellar medium) of an off-axis relativistic jet.

The large amount of lanthanide-rich ejecta inferred from kilonova observations ($\gtrsim 10^{-2}\,\mathrm{M_\odot}$, e.g.~\citealt{Shibata2017,Nicholl2017,Cowperthwaite2017,Villar2017,Perego2017a}) has been interpreted \citep{Margalit2017} as being the result of strong  winds from the accretion disc around the merger remnant. Since these winds are expected to unbind a few tens of percent of the accretion disc mass \citep[e.g.][]{Siegel2017}, this in turn requires the disc to be rather massive, of the order of $M_\mathrm{disc}\sim 10^{-1}\,\mathrm{M_\odot}$ \citep{Radice2019}. 

The kinetic energy in the GRB 170817A jet has been constrained by several groups, based on multi-wavelength modelling of the non-thermal afterglow and on the VLBI centroid motion (see Sect. \ref{sec:jet_energy} and Fig.~\ref{fig:jet_energy_posterior}). All estimates essentially agree within the uncertainties, clustering slightly below $E_\mathrm{K,jet}\sim 10^{50}\,\mathrm{erg}$.

We therefore have, for the first time, two independent estimates of the energy in a GRB jet and of the mass of the accretion disc around the compact object that produced it, which enables us to estimate the accretion-to-jet energy conversion efficiency. By simply using the gross estimates above, one finds that this efficiency is of the order of $\eta\sim E_\mathrm{K,jet}/M_\mathrm{disc}c^2\sim 10^{-3}$. While this number may appear surprisingly low, this level of efficiency was anticipated in previous studies that attempted to connect the energy in SGRB jets with the underlying disc masses \citep[e.g.][]{Giacomazzo2013,Ascenzi2019,Barbieri2019}.

In this work, we assume the remnant of GW170817 to be a BH surrounded by an accretion disc\footnote{Jet launching is still possible \citep[e.g.][]{Thompson1994,Bucciantini2008,Metzger2011,Mosta2020} in the case of a long-lived neutron star remnant, though it seems disfavored by recent simulations \citep[e.g.][]{Ciolfi2020b}. In this case, the launch of the jet would be powered by the rotation of the magnetised neutron star remnant, so that our definition of efficiency would not be applicable.}; the actual nature of the remnant could not be identified based on GW observations alone (\citealt{LVC2017postmerger,LVC2019postmerger}), but the collapse to a BH after a short hyper-massive neutron star (HMNS) phase seems the most likely outcome for this system (see for example \citealt{Gill2019} for a thorough discussion).\ We carefully constructed a posterior probability distribution for this efficiency in order to account for the (large) uncertainties in both the disc mass and the jet energy, and we compared the result with expectations based on the two main candidate jet-launching processes, namely the \citet{Blandford1977} and the neutrino-antineutrino annihilation mechanisms \citep{Eichler1989,Meszaros1992}. We show that such a low efficiency is expected in SGRBs for both mechanisms and (unfortunately) cannot be used to distinguish between the two.

\section{Disc mass in GW170817}\label{sec:disc_mass}
Several mechanisms are thought to cause dynamical mass ejection during the merger of two neutron stars  \citep{Shibata2019,Radice2018,Bauswein2013,Hotokezaka2013,Rosswog1999,Davies1994}, such as~tidal interactions between the two stars, shocks that form as a consequence of the collision, and violent oscillations of the highly oblate remnant shortly after the two stars have merged. 
Numerical simulations show that such matter is typically ejected with a broad range of (mostly non-relativistic) velocities and internal energies, but a large fraction generally remain gravitationally bound and form an accretion disc around the merger remnant. The amount of disc mass ultimately depends on the intrinsic properties of the binary prior to the merger, namely~the equation of state (EoS) of neutron star matter, the component masses, spins, orbit eccentricity, and possibly the magnetic fields. The pre-merger magnetic field hardly affects the dynamics of the inspiral\footnote{Unless it is of extreme intensity, $B\gtrsim 10^{17}\,\mathrm{G}$ -- \citealt{Giacomazzo2009} -- but it is unclear how such a high magnetic field could survive during the neutron star lifetime before the merger}.
On the other hand, the magnetic field intensity in the post-merger can influence the lifetime of a meta-stable HMNS remnant \citep{Giacomazzo2011} and the intensity of its winds \citep[e.g.][]{Ciolfi2020,Ciolfi2020c,Mosta2020}, which can in turn affect the disc mass before the HMNS collapses to a BH. Nevertheless, the magnetic field amplification by Kelvin-Helmholtz instabilities during the merger \citep{Kiuchi2014,Kiuchi2018} likely erases any memory of the initial magnetic field; therefore, we neglect here any dependence on the pre-merger magnetic field for the sake of simplicity as it is, in any case, most likely well below our uncertainties. While their effect on the dynamics and mass ejection could be relevant \citep{East2019}, spins are usually expected to be low at merger, based on observations of Galactic double neutron star systems \citep{Lorimer2008} and their decay due to spin-down prior to merger \citep{Stovall2018}; we will therefore assume spins to be negligible in this work. The orbit eccentricity prior to the merger is expected to be low in most realistic cases (\citealt{Kowalska2011}) unless the merger happens soon after a dynamical interaction in a dense stellar environment, though the associated rates are expected to be low (e.g.~\citealt{Ye2020}) due to circularisation caused by GW emission. We are therefore left with the neutron star masses and the EoS as the only parameters upon which the disc mass can depend. 

\subsection{Disc mass posterior probability}

\begin{figure} 
\centering
 \includegraphics[width=\columnwidth]{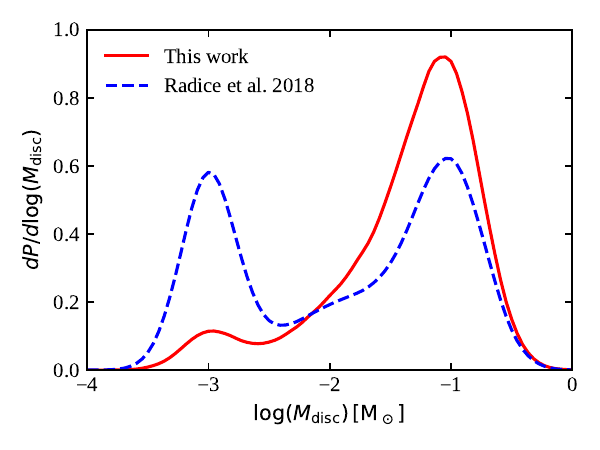}
 \caption{GW170817 accretion disc mass posterior distributions. The solid red line shows the posterior probability distribution of the logarithm of the accretion disc mass (in solar masses) for the low-spin LVC priors. The dashed blue line shows the corresponding result that would have been obtained using the disc mass fitting formula from \citet{Radice2018}. }
 \label{fig:disc_mass_posterior}
\end{figure}

Based on a large suite of general relativistic hydrodynamical simulations, \citet[][R18 hereafter]{Radice2018} devised a fitting formula for the disc mass that depends solely on the binary effective dimensionless tidal deformability $\tilde \Lambda$, which is a combination of the masses and tidal deformabilities (which are in turn determined by the EoS) of the binary components. This quantity is of particular interest because it appears in the leading post-Newtonian term that describes tidal effects on the GW waveform \citep{Flanagan2008} and is therefore the best constrained combination of EoS-related parameters from GW analysis. The fitting formula is accurate to $\sim 50\%$ when compared to most of the simulations in the suite of R18, but these are based on a limited number of different EoSs, and, most importantly, they do not comprise significantly unequal-mass binaries\footnote{These simulations also do not include magnetic fields, which could affect the disc mass around the BH remnant by their effects on pressure and viscosity, especially during a possible HMNS transient phase. This is mitigated, though, by their implementation of an effective `large-eddy' simulation method \citep{Radice2017} that partly reproduces the effects of small-scale magneto-hydrodynamical turbulence. Such a method requires fixing an effective mixing length parameter: The \citet{Barbieri2020} fitting formula was calibrated on all their models, which include a range of mixing lengths, in order to effectively account for the systematic uncertainty that stems from the uncertainty on this parameter.}. \citet{Kiuchi2019} later showed that, at fixed $\tilde \Lambda$, unequal-mass systems can yield a larger disc mass due to the increased tidal deformation of the lighter star. \citet{Barbieri2020} presented a new, simple fitting formula based on a toy model of the mass ejection in a neutron star binary merger. The formula depends on $\tilde \Lambda$ and on the masses of the primary ($M_1$) and secondary ($M_2$) components of the binary. After fitting the free parameters to the results of numerical simulations collected from R18, \citet{Kiuchi2019}, \citet{Vincent2019}, and \citet{Bernuzzi2020}, they find that such a formula predicts the correct disc mass for both equal- and unequal-mass systems, with an error that is comparable to that of the original R18 formula for equal-mass binaries. We applied this formula\footnote{During the preparation of this work, two different pre-prints with two alternative fitting formulae for the disc mass were circulated \citep{Kruger2020,Dietrich2020}. We verified that our conclusions remained unchanged when employing either of these alternative fitting formulae to compute the disc mass.} to the posterior samples from the GW parameter estimation of GW170817 \citep{Abbott2019} for low-spin priors (following our assumption of negligible spins). Following \citet{Radice2019}, we accounted for the uncertainty in the disc mass fitting formula as follows: For each GW posterior sample, we computed the disc mass using the fitting formula from \citet{Barbieri2020}, and then we extracted 100 samples from a log-normal distribution centred at that value, with a dispersion $\sigma=0.5\,\mathrm{M_\odot}$. The resulting disc mass posterior distributions are shown in Fig.~\ref{fig:disc_mass_posterior}. The most probable disc mass value is around $M_\mathrm{disc}\sim 0.1\,\mathrm{M_\odot}$, with a tail of smaller probability that extends down to $\sim 10^{-3}\,\mathrm{M_\odot}$. Compared to the predictions obtained using the formula from \citet{Radice2018} (dashed blue line), the bimodality in the predicted disc masses is essentially suppressed, in agreement with the argument by \citet{Kiuchi2019}. This indicates that the most compact configurations compatible with GW170817 -- those with $\tilde \Lambda \lesssim 400$, which produce the low-disc-mass peak in the \citealt{Radice2018} distribution -- also have consistently lower mass ratios $q=M_2/M_1$ on average.

\subsection{Accreted disc mass}
Long-term numerical simulations of merger remnant accretion discs \citep[e.g.][]{Fujibayashi2020,Christie2019,Fernandez2019,Siegel2018,Siegel2017,Just2015,Fernandez2013} indicate that a significant fraction ($f_\mathrm{w}=0.1$ to $0.5$) of the disc mass can be lost in the form of winds in these systems, therefore lowering the actual accretion rate and the final accreted mass. The majority of such mass loss takes place as the disc spreads viscously and transitions to the advection-dominated accretion flow (ADAF) phase, during which it inflates due to viscous heating and convective motions \citep{Metzger2009,Fernandez2013,Just2015}. Observations of AT2017gfo, the kilonova associated with GW170817, and the subsequent modelling indeed seem to indicate that a significant fraction of the kilonova ejecta mass did originate in disc winds (\citealt{Margalit2017}, see the introduction for additional references). We therefore defined $M_\mathrm{a}=(1-f_\mathrm{w})M_\mathrm{disc}$ as the accreted mass, and we adopted the fiducial value $f_\mathrm{w}=0.3$. The exact value of this parameter does not significantly affect our conclusions.

\section{Jet energy in GRB~170817A}\label{sec:jet_energy}
Inferring the true energy of a GRB jet is not straightforward in general, even in cases where both the prompt and afterglow emissions have been extensively observed. The understanding of the prompt emission phase is hampered by theoretical uncertainties on several aspects, regarding both the dominant form of energy in the jet (either magnetic or bulk kinetic) and the way this energy is transformed into the radiation we observe. Moreover, during the prompt emission, the emitting material is thought to be in highly relativistic motion, which essentially renders it impossible to infer the jet opening angle -- and therefore derive the jet true energy -- from observations of this emission phase alone. 

Observations of the afterglow provide, in principle, a better tool for inferring the true jet energy. As the jet material collides with the interstellar medium (ISM), it drives a shock that heats up particles, which can then radiate \citep{Paczynski1993,Meszaros1997}. Moreover, as the shock sweeps the ISM, the shocked fluid slows down and relativistic beaming is reduced, thus allowing the observer to infer the actual jet opening angle \citep{Rhoads1997}. 

\begin{figure}
\centering
 \includegraphics[width=\columnwidth]{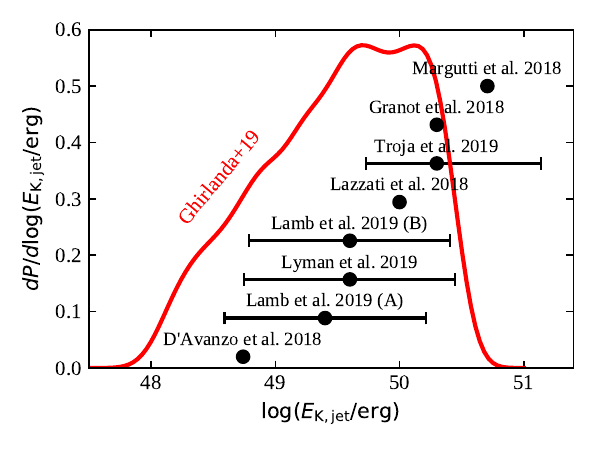}
 \caption{Kinetic energy in the GRB 1790817A jet during the afterglow phase. The solid red line shows the posterior distribution constructed by fitting a structured jet afterglow model to the multi-wavelength afterglow and VLBI centroid motion by \cite{Ghirlanda2019}. Black dots with error bars represent the estimates presented in several other papers, all based on afterglow fitting. All estimates essentially agree within the uncertainties. } 
 \label{fig:jet_energy_posterior}
\end{figure}

In the case of GRB 170817A, the jet kinetic energy has been estimated by several groups based on the multi-wavelength modelling of the afterglow emission. Despite the modelling uncertainties, the various estimates (see Fig.~\ref{fig:jet_energy_posterior}, where we show estimates from \citealt{Ghirlanda2019,Lamb2019,Lyman2018,Troja2019,Lazzati2018,Margutti2018,DAvanzo2018,Granot2018}) essentially agree within their uncertainties, clustering around $E_\mathrm{K,jet}\sim 10^{50}\,\mathrm{erg}$ (see also the recent pre-print by \citealt{Lamb2020}, who find a similar energy under significantly different assumptions). This is the total kinetic energy contained in the jet material that caused the relativistic shock that produced the observed (X-ray, optical, and radio) afterglow of GRB 170817A. This is not, in principle, the same as the jet energy actually produced by the merger remnant since a fraction, $f_\mathrm{bk}$, is spent to break out of the kilonova ejecta and another fraction, $f_\gamma$, is lost in the prompt emission. \citet{Duffell2018} has shown, though, that the jet energy spent in breaking out of the ejecta is roughly $E_\mathrm{bk}=0.05 \theta_\mathrm{j}^2 E_\mathrm{ej}$, where $\theta_\mathrm{j}$ is the opening angle of the jet at launch and $E_\mathrm{ej}$ is the total kinetic energy in the ejecta. Even assuming ejecta as massive as $M_\mathrm{ej}= 0.1\,\mathrm{M_\odot}$ with an average velocity $v_\mathrm{ej}\sim 0.1 c$ and a relatively large jet opening angle at launch $\theta_\mathrm{j}=0.3\,\mathrm{rad}$ (the jet is then collimated by the ejecta prior to breakout), the energy $E_\mathrm{bk}$ barely reaches $5\times 10^{48}\,\mathrm{erg}$ and can therefore be safely neglected (i.e.~we set $f_\mathrm{bk}=0$). As for $f_\gamma$, this value can generally be estimated by comparing the jet kinetic energy inferred from early X-ray afterglow observations to the energy radiated in the prompt emission\footnote{This approach may be questioned since there are indications \citep[e.g.][]{Ghisellini2007,D'Avanzo2012} that the early X-ray afterglow of a relevant fraction of GRBs is dominated by a component that is  linked to central engine activity. Nevertheless, focusing on late-time X-ray observations, \citet{D'Avanzo2012} still find typical efficiencies below 10 percent. See also \citet{Nemmen2012}, who find indications that 15 percent is a typical radiative efficiency in all astrophysical jets.} \citep[e.g.][]{Zhang2007}. This leads to a variety of values for the efficiency, which in some cases are as high as 90\%. Recently, though, \citet{Beniamini2015} and \citet{Beniamini2016} found that these values were typically overestimated as Compton cooling was overlooked in the modelling of these afterglows in part due to a bias in the adopted post-shock magnetic field equipartition parameter $\epsilon_B$. Removing this bias, they showed that the typical prompt emission efficiency in GRBs is $f_\mathrm{\gamma}\sim0.15$, which is the value we adopt here. Therefore, our estimate of the actual jet energy produced by the merger remnant is $E_\mathrm{j}=2\times E_\mathrm{K,jet}/(1-f_\mathrm{\gamma})\approx 2.36 E_\mathrm{K,jet}$ (here the factor of $2$ is inserted to account for the counterjet, which is assumed to be identical to the jet that produced the observed emission), but we note that our uncertainties make us essentially insensitive to the precise value of $f_\mathrm{\gamma}$ as long as it is significantly less than unity. 

For our fiducial estimate of $E_\mathrm{K,jet}$, we used the posterior from \citet{Ghirlanda2019}, which accounts for the uncertainty on all parameters and is based on multi-wavelength afterglow fitting, including the very-long-baseline interferometry (VLBI) centroid motion. The posterior is shown by the red line in Fig.~\ref{fig:jet_energy_posterior}, which compares it to several estimates from other authors.

\section{Accretion-to-jet energy conversion efficiency in GW170817}

\begin{figure}
\centering
 \includegraphics[width=\columnwidth]{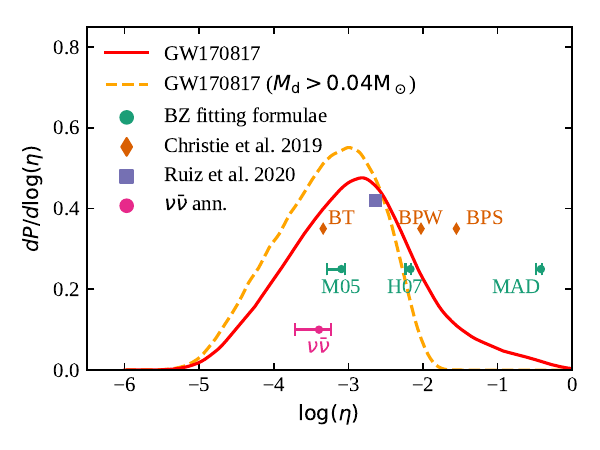}
 \caption{Posterior probability density distribution of the accretion-to-jet energy conversion efficiency $\eta=E_\mathrm{j}/M_\mathrm{a}c^2$ in GW170817 (solid red line), assuming $f_\mathrm{w}=0.3$ and $f_\mathrm{\gamma}=0.15$. The dashed orange line shows the same quantity, but with the disc mass in GW170817 required to be equal or greater than $0.04\,\mathrm{M_\odot}$ \citep{Radice2019}. Black error bars show the expected efficiency for the \citet{Blandford1977} and $\nu\bar\nu$ annihilation mechanisms, according to different prescriptions (see text).  } 
 \label{fig:efficiency}
\end{figure}

Combining the posterior distributions of the accreted disc mass $M_\mathrm{a}=(1-f_\mathrm{w}) M_\mathrm{disc}$ (Sect. \ref{sec:disc_mass}) and the posterior distributions of the total jet energy $E_\mathrm{j}=2\times E_\mathrm{K,jet}/(1-f_\mathrm{\gamma})$ (Sect. \ref{sec:jet_energy}), we can derive the posterior distribution of the accretion-to-jet energy conversion efficiency $\eta = E_\mathrm{j}/M_\mathrm{a}c^2=2 E_\mathrm{K,jet}/(1-f_\mathrm{\gamma})(1-f_\mathrm{w})M_\mathrm{disc}c^2$. Figure~\ref{fig:efficiency} shows the resulting posterior probability distribution for $\eta$ (solid red line), assuming $f_\mathrm{w}=0.3$ and $f_\gamma=0.15$, as discussed in the previous sections. The 1 sigma confidence interval is $\eta=1.2_{-1.0}^{+7.0}\times 10^{-3}$. If we require the GW170817 disc mass to be $M_\mathrm{disc}>0.04\,\mathrm{M_\odot}$, based on the constraint from kilonova observations \citep{Radice2019,Margalit2017}, the posterior (dashed orange line) is pushed further towards low efficiencies; the 1 sigma confidence interval in this case is $\eta=0.6_{-0.5}^{+2.0}\times 10^{-3}$. These efficiencies are summarised, along with theoretically expected values (discussed in the following sections), in Table~\ref{tab:efficiencies}.

\section{Comparison with theoretical expectations}

It is informative to compare the GW170817 accretion-to-jet energy conversion efficiency estimate derived in the previous section with theoretical expectations based on the two main jet launching mechanism candidates, namely the \citet{Blandford1977} magneto-hydrodynamical (MHD) mechanism and the neutrino energy deposition mechanism \citep{Eichler1989}. 

\subsection{Blandford-Znajek mechanism}
\label{sec:BZ_eff}

\begin{figure}
 \includegraphics[width=\columnwidth]{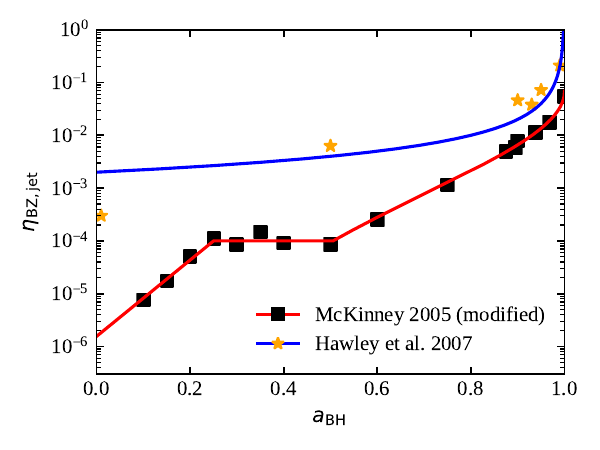}
 \caption{Accretion-to-jet energy conversion efficiency of the \citet{Blandford1977} mechanism as a function of the BH spin parameter $a_\mathrm{BH}$ from the GRMHD simulations of \citet[][black squares]{McKinney2005} and \citet[][orange stars]{Hawley2007}. The red and blue curves are fitting functions Eqs.~\ref{eq:BZ_eff_M05} and \ref{eq:BZ_eff_H07}, respectively.}
 \label{fig:BZ_eff} 
\end{figure}

The \citet[][BZ hereafter]{Blandford1977} mechanism can produce efficient energy extraction from a BH threaded by a large-scale magnetic field in relative rotation with respect to the BH. The mechanism operates in force-free regions (i.e.~regions where the matter contribution to the stress energy tensor is negligible with respect to the electromagnetic contribution) close to the BH horizon \citep[e.g.][]{Komissarov2004,McKinney2004,McKinney2005,DeVilliers2005,Hawley2006,Hawley2007,Tchekhovskoy2010}. In the case of a Kerr BH with dimensionless spin parameter $a_\mathrm{BH}$, the extracted power depends on the strength of the radial component $B^r$ of the magnetic field at the horizon, as well as on the spin parameter \citep{Blandford1977,Komissarov2004,Tchekhovskoy2010}. The magnetic field is most naturally brought to the BH horizon by an accretion disc, where it can be amplified by means of the magneto-rotational instability \citep[MRI,][]{Balbus1991,Hawley1991}. In stationary accretion conditions, it is plausible to expect the MRI to saturate to a definite disc magnetisation that does not depend on the seed magnetic field strength. Indeed, based on a series of axisymmetric general-relativistic magneto-hydrodynamical (GRMHD) simulations of Kerr BHs surrounded by thick accretion discs seeded with a dipole magnetic field (described in \citealt{McKinney2004}), \citet{McKinney2005} found that, after an initial transient phase, the disc magnetisation saturates and the accretion-to-jet energy conversion efficiency of the BZ process stabilises depending solely on the BH spin, which is well described by the simple relation $\eta_\mathrm{BZ,jet}\sim 0.068\,\Omega_\mathrm{H}^5(a_\mathrm{BH})$ for $a>0.5$. Here, $\Omega_\mathrm{H}=a_\mathrm{BH}/(1+\sqrt{1-a_\mathrm{BH}^2})$ is the dimensionless angular frequency at the BH horizon. The full set of their simulations (see Fig.~\ref{fig:BZ_eff}), including those with $a<0.5$, can be fitted by the function 
\begin{equation}
 \eta_\mathrm{BZ,M05}=\left\lbrace\begin{array}{lr}
                                                  10^{-4}\,\mathrm{e}^{(a_\mathrm{BH}-0.25)/0.06} & a_\mathrm{BH}\leq 0.25 \\
                                                  10^{-4} & 0.25<a_\mathrm{BH}\leq 0.505 \\
                                                  0.068\,\Omega_\mathrm{H}^5 & a_\mathrm{BH}>0.505 \\
                                                 \end{array}
\right.
\label{eq:BZ_eff_M05}
.\end{equation}
Three-dimensional simulations by a different group, described in \citet{DeVilliers2005}, \citet{Hawley2006}, and \citet{Hawley2007}, find a somewhat different saturation efficiency, which can be described (see again Fig.~\ref{fig:BZ_eff}) by the simple expression
\begin{equation}
 \eta_\mathrm{BZ,H07}=\frac{0.002}{1-a_\mathrm{BH}}
\label{eq:BZ_eff_H07}
.\end{equation}
This efficiency is obtained in simulations where the seed magnetic field is poloidal, and the authors note that the efficiency of jet launching critically depends on the seed magnetic field configuration, with purely toroidal initial fields leading to weak or absent jets. On the other hand, \citet{Liska2020} more recently found that even an initially purely toroidal magnetic field can lead to a strong jet, provided that the simulation is run for long enough (several$\times 10^{4}$ dynamical times) for an MHD dynamo-like process \citep{Moffatt1978} to convert part of the magnetic field into a large-scale poloidal component (see the discussion in \citealt{Liska2020}). In other words, the expectation that the disc magnetisation eventually saturates regardless of its initial strength and configuration seems fulfilled, provided that the accretion lasts long enough and that the accretion conditions remain otherwise unchanged in the meantime.

The above arguments assume the seed magnetic field to be initially sub-dominant. At the other extreme end, the highest BZ efficiency is reached \citep[e.g.][]{Narayan2003,Tchekhovskoy2011,McKinney2012} when accretion takes the form of a magnetically arrested disc \citep[MAD,][]{Bisnovatyi-Kogan1976}. In this case, the magnetic field energy density is so high that it dominates the disc dynamics, and the magnetic flux through the BH horizon is maximised, leading to efficiencies that can exceed 100\% (we note that the ultimate energy source of the process is BH rotation rather than accretion power). The jet efficiency can also be expressed in this case as a function of the BH spin \citep{Tchekhovskoy2015}, namely\footnote{We omit here, for the sake of simplicity, a linear scaling with the height-to-radius (H/R) ratio,
which is set to 0.2 as in \citet{Tchekhovskoy2015}, as this does not affect our conclusions.}
\begin{equation}
 \eta_\mathrm{BZ,MAD}=3\Omega_\mathrm{H}^3(1-0.38\Omega_\mathrm{H})^2(1+0.35\Omega_\mathrm{H}-0.58\Omega_\mathrm{H}^2)
 \label{eq:BZ_eff_MAD}
.\end{equation}

\subsubsection{Expected Blandford-Znajek efficiency in GW170817}
A first simple (but naive -- see next section) estimate of the expected BZ efficiency in GW170817 can be obtained by evaluating the fitting formulae from the previous section at the GW170817 BH remnant spin (computed as described in Appendix~\ref{sec:BH_mass_and_spin}). Doing so, we obtained the theoretical efficiency estimates shown in Fig.~\ref{fig:efficiency}, namely $\eta_\mathrm{BZ,M05}=8.0^{+0.9}_{-3.0}\times 10^{-4}$, $\eta_\mathrm{BZ,H07}=6.8^{+0.2}_{-1.0}\times 10^{-3}$, and $\eta_\mathrm{BZ,MAD}=0.38^{+0.01}_{-0.05}$. While the MAD case is clearly excluded, the estimates based on \citet{McKinney2005} and \citet{Hawley2007} are both in good agreement with our `measured' efficiency in GW170817, given the large uncertainties. If we assume the disc mass constraint $M_\mathrm{disc}>0.04\,\mathrm{M_\odot}$ \citep{Radice2019,Margalit2017}, then the efficiency from \citet{Hawley2007} falls outside of the 1 sigma confidence interval.

\subsubsection{Efficiency found in GRMHD simulations of generic BNS mergers}
One thing to keep in mind when considering the efficiencies calculated in the previous section is that the simulations on which they are based are actually more relevant to BH-accretion disc systems hosted by active galactic nuclei, in which accretion is thought to take place on sufficiently long time scales to settle on a definite state with a constant accretion rate (on the time scales of interest). Accretion on the remnant of a BNS merger, on the other hand, is essentially transient, with a declining accretion rate (see Sect. \ref{sec:nunu}).  Under typical circumstances, it most likely evolves over four stages \citep[e.g.][]{Just2015,Christie2019}: (1) Initially, the accretion rate is high enough for the disc to be optically thick to neutrinos, so that a strong and fast neutrino-driven wind is produced by the inner part of the disc. During this short phase, jet launching is likely hampered by baryon pollution in the funnel above the BH. (2) As the accretion rate decreases, the disc transitions to a neutrino-dominated accretion flow (NDAF) state, where it efficiently cools by neutrino emission, and the density in the polar region drops, allowing for the formation of a force-free region. This is a requirement for the BZ process to take place. (3) The decreasing accretion rate eventually renders neutrino cooling inefficient again, leading to an ADAF phase during which viscous heating inflates and expands the the disc, leading to strong disc winds. (4) Finally, the accretion rate becomes low enough for the magnetic field energy density to overcome that of the accreting matter, leading to an MAD \citep{Tchekhovskoy2015}. During this evolution, the BZ efficiency increases \citep{Christie2019}, but this is compensated for by the decreasing accretion rate so that, in the end, the average efficiency typically falls well below the MAD case. 

A recent well-suited example is the set of high-resolution, long-term GRMHD simulations, described in \citet{Christie2019}, of a BH surrounded by an accretion disc with initial conditions that represent those that immediately follow a BNS merger, with a BH mass $M_\mathrm{BH}=3\,\mathrm{M_\odot}$,  a spin parameter $a_\mathrm{BH}=0.8$, and a disc (torus) of mass $M_\mathrm{disc}=0.033\,\mathrm{M_\odot}$. The simulations were designed to investigate the impact of magnetic field geometry on the properties of the outflows (both relativistic and non-relativistic) produced by such a system. Two simulations, called BPS and BPW, were initialised with a poloidal magnetic field within the disc, differing only in the degree of magnetisation (BPS had a stronger magnetisation than BPW). The third simulation, BT,
was seeded with a toroidal magnetic field within the disc\footnote{A predominantly toroidal configuration is expected in neutron star mergers, due to the stretching of neutron star material undergoing tidal disruption, and to flux freezing, see e.g.~\citealt{Kiuchi2014,Kawamura2016}.}
In all cases, the system was able to launch a relativistic jet, but with differing efficiencies:  The BPS simulation produced bipolar jets with $E_\mathrm{j}\sim 10^{51}\,\mathrm{erg}$, and the fraction of disc mass lost in winds was $f_\mathrm{w}=0.4$, leading to $\eta_\mathrm{BZ}\sim 2.8 \times 10^{-2}$. For the BPW simulation, $E_\mathrm{j}\sim 3.9\times 10^{50}\,\mathrm{erg}$ and $f_\mathrm{w}=0.3$, yielding $\eta_\mathrm{BZ}\sim 9.4 \times 10^{-3}$. Finally, simulation BT resulted in significantly weaker jets with $E_\mathrm{j}\sim 2\times 10^{49}\,\mathrm{erg}$. Strong disc winds were still present, with $f_\mathrm{w}=0.27$, so that $\eta_\mathrm{BZ}\sim 4.6 \times 10^{-4}$. 
Our maximum posterior estimate of the efficiency falls in between the BT and BPW cases, which seems reasonable since the expected magnetic field configuration right after the merger is predominantly toroidal, but a weak poloidal component is expected.

Another set of recent GRMHD simulations of BNS mergers that resolve the relativistic jet launching by the \citet{Blandford1977} mechanism in the post-merger phase are those described in \citet{Ruiz2019} and \citet{Ruiz2020}. These simulations start a few orbits before the merger, so that, in this case, the magnetic field configuration in the torus is self-consistently determined by the merger dynamics\footnote{We note, though, that the resolution is not high enough to resolve the magnetic field amplification by the Kelvin-Helmholtz instability; for this reason, the neutron stars are endowed with strong magnetic fields $B\gtrsim 10^{15}\,\mathrm{G}$ before the merger.}
Essentially, regardless of the initial configuration, the authors find BZ efficiencies\footnote{These values correspond to a slightly different definition of efficiency, i.e.~the ratio between accretion rate and jet luminosity. Since these simulations are not run until the end of accretion, our definition is not applicable.} $\eta_\mathrm{BZ}\sim 2-3\times 10^{-3}$ in all jet-producing simulations, in good agreement with our results. 

We therefore conclude that the accretion-to-jet energy conversion efficiency in GW170817 is consistent with theoretical expectations for the BZ mechanism in the presence of an initial magnetic field whose energy density does not dominate over that of the accreting matter (i.e.~the disc is not in the MAD state) during most of the accretion, and whose configuration is predominantly toroidal right after the merger.

\subsection{Neutrino mechanism}
\label{sec:nunu}

While the \citet{Blandford1977} process today is widely regarded as the most likely jet-launching mechanism in GRBs, the first works to propose neutron star mergers as potential progenitors of GRBs \citep[][]{Eichler1989,Meszaros1992,Mochkovitch1993,Mochkovitch1995} actually envisioned energy deposition by the annihilation of neutrino-antineutrino pairs in the vicinity of the merger remnant (`$\nu\bar\nu$ mechanism' hereafter) as the process responsible for powering the jet. Whether such a scenario can realistically power GRB jets (including long GRBs, e.g.~\citealt{Popham1999,Kohri2005,Lee2006}) has long been a subject of debate.   Several works have studied the detailed structure and stability of accretion discs in NDAF conditions \citep[e.g.][]{Popham1999,Asano2001,Narayan2001,DiMatteo2002,Kohri2002,Kohri2005,Lee2006,Chen2007,Birkl2007,Kawanaka2007,Janiuk2007,Rossi2007,Zhang2009,Janiuk2010,Kawanaka2012,Pan2012,Janiuk2013,Kawanaka2013,Kawanaka2013b,Liu2014,Liu2015,Liu2015a,Janiuk2017,Liu2018,Kawanaka2019}; others attempted to directly simulate the neutron star merger (or its remnant) and the formation of a jet by the $\nu\bar\nu$ mechanism \citep[e.g.][]{Ruffert1997,Ruffert1999,Rosswog2003,Shibata2007,Dessart2009,Just2015,Just2016,Perego2017,Siegel2017}. While most works on the former aspect find promising neutrino luminosities and energy deposition rates, the latter investigations mostly indicate that the $\nu\bar\nu$ mechanism faces some apparently serious difficulties. In particular \citep{Just2016}, the typical BNS post-merger environment could be sufficiently dense as to prevent a $\nu\bar\nu$-powered jet to successfully break out, in part due to the short duration of the high neutrino luminosity phase.
\begin{figure} 
 \includegraphics[width=\columnwidth]{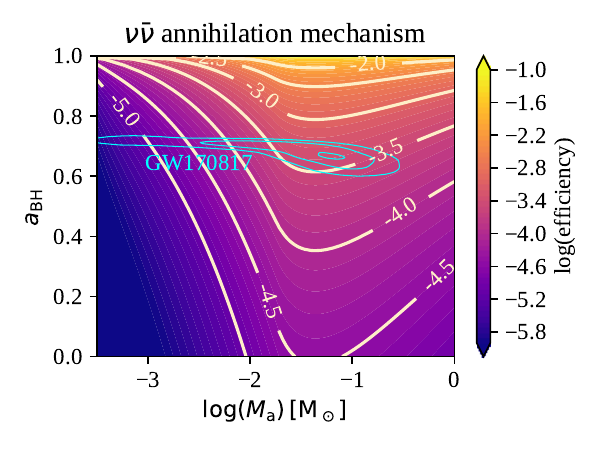}
 \caption{Accretion-to-jet energy conversion efficiency of the $\nu\bar\nu$ mechanism as a function of the accreted mass $M_\mathrm{a}$ and BH spin $a_\mathrm{BH}$, assuming a $2.6\,M_\odot$ BH, $\alpha=2$, and $t_0=10\,\mathrm{ms}$. The three overlaid cyan contours contain, respectively, $10\%$, $68\%$, and $95\%$ of the joint GW170817 remnant BH spin -- accreted mass posterior probability density, assuming $f_\mathrm{w}=0.3$.}
 \label{fig:nu_nu_eff} 
\end{figure}

In what follows, we aim to test whether, in principle, the $\nu\bar\nu$ mechanism accretion-to-jet energy conversion efficiency is compatible with our measured 
one. In order to estimate the expected efficiency, we adopted the parameterisation of the jet power given in \citet{Leng2014}, which relies on energy deposition rates computed by \cite{Zalamea2011} based on the relativistic model of a neutrino-cooled accretion disc around a Kerr BH described in \citet{Chen2007}, as well as on general-relativistic ray tracing of the neutrino trajectories. 

We assume the disc accretion rate to evolve with time as a power law, namely
\begin{equation}
 \dot M (t) = (\alpha-1)\frac{M_\mathrm{a}}{t_0}\left(\frac{t}{t_0}\right)^{-\alpha}
 \label{eq:Mdot}
\end{equation}
for $t>t_0$, while $\dot M = 0$ for $t<t_0$. When $1.5\lesssim \alpha \lesssim 2.5$, this simple prescription is a fair description of the actual output of detailed numerical simulations \citep[e.g.][]{Fernandez2019,Just2016}. Time here is measured from the formation of the central BH, and $t_0$ is the time to the onset of accretion, which is of the order of a few dynamical times. 
This allows us, following \citet{Leng2014}, to write the jet power as a function of the accretion rate $\dot M$ and BH spin parameter $a_\mathrm{BH}$ as follows
\begin{equation}
\dot E_{\nu\bar\nu} \sim \dot E_\mathrm{sat}\left\lbrace\begin{array}{lr}
                                              1 & \dot M \geq \dot M_\mathrm{sat}\\
                                              \left(\frac{\dot M}{\dot M_\mathrm{sat}}\right)^{9/4} & \dot M_\mathrm{ign} < \dot M < \dot M_\mathrm{sat}\\
                                              0 & \dot M \leq \dot M_\mathrm{ign}\\
                                             \end{array}\right.
\label{eq:Zalamea}
, \end{equation}
where
\begin{equation}
 \dot E_\mathrm{sat} = 5.5 \times 10^{52} \left(\frac{R_\mathrm{ISCO}}{2 R_\mathrm{g}}\right)^{-4.8}\left(\frac{M_\mathrm{BH}}{2.5 M_\odot}\right)^{-3/2} \mathrm{erg\,s^{-1}}
.\end{equation}
Here, $R_\mathrm{g}=G M_\mathrm{BH}/c^2$ is the gravitational radius, $M_\mathrm{BH}$ is the central BH mass, $R_\mathrm{ISCO}= R_\mathrm{ISCO}(a_\mathrm{BH})$ is the innermost stable circular orbit, $M_\mathrm{sat}=1.8 M_\odot\,\mathrm{s^{-1}}$ is the saturation accretion rate (above which the jet power is assumed to saturate), and $\dot M_\mathrm{ign}=0.021 M_\odot\,\mathrm{s^{-1}}$ is the accretion rate below which neutrino cooling becomes inefficient.

Assuming this description holds true at all times during the evolution of $\dot M(t),$ and neglecting the small change (in the SGRB case) in the BH mass and spin due to accretion (and jet production), the jet energy can be obtained by integrating Eq.~\ref{eq:Zalamea} over time analytically. For that purpose, it is useful to identify two dimensionless transition times: 
\begin{equation}
 \tau_\mathrm{sat}= \left(\frac{(\alpha-1) M_\mathrm{a}}{ t_0 \dot M_\mathrm{sat}}\right)^{1/\alpha}
\label{eq:t_sat}
\end{equation}
and 
\begin{equation}
 \tau_\mathrm{ign}= \left(\frac{(\alpha-1) M_\mathrm{a}}{ t_0 \dot M_\mathrm{ign}}\right)^{1/\alpha}
\label{eq:t_ign}
.\end{equation}
Setting $\delta=(9/4)\alpha -1$, one can then write the final jet energy produced by neutrino-antineutrino annihilation as 
\begin{equation}
 E_\mathrm{j,\nu\bar\nu} \sim \dot E_\mathrm{sat}t_0 \tau_\mathrm{sat}\left[\max\left(1-\tau_\mathrm{sat}^{-1},0\right) + \frac{1}{\delta}\left(1-\left(\frac{\tau_\mathrm{sat}}{\tau_\mathrm{ign}}\right)^{\delta}\right) \right]
.\end{equation}The ratio $\eta_\mathrm{\nu\bar\nu}(a_\mathrm{BH},M_\mathrm{a},M_\mathrm{BH},\alpha,t_0) = E_\mathrm{j,\nu\bar\nu}/M_a c^2$ then represents the accretion-to-jet energy conversion efficiency of the neutrino-antineutrino annihilation mechanism. 
Figure~\ref{fig:nu_nu_eff} shows the resulting efficiency contours on the $(M_\mathrm{a},a_\mathrm{BH})$ plane, assuming $\alpha=2$, $t_0=10\,\mathrm{ms}$, and  $M_\mathrm{BH}=2.6\,\mathrm{M_\odot}$. Overlaid cyan contours show the joint posterior probability density $P(M_\mathrm{a},a_\mathrm{BH})$ for GW170817 assuming $f_\mathrm{w}=0.3$, obtained after marginalising over the BH mass (see Appendix \ref{sec:BH_mass_and_spin} for the computation of the remnant BH mass and spin). The corresponding prediction for the neutrino mechanism efficiency in GW170817, adopting uniform  priors on $\alpha$ in the range (1.5, 2.5) and on $t_0/\mathrm{ms}$ in the range (5, 30), amounts to $\eta_{\nu\bar\nu}=4.0^{+1.8}_{-2.1}\times 10^{-4}$, which is shown in Fig.~\ref{fig:efficiency}. This efficiency is only slightly lower than that of the BZ mechanism as predicted by the \citet{McKinney2005} fitting formula, and it is compatible with the measured one (see Table~\ref{tab:efficiencies}). We therefore cannot exclude, on the basis of our determination of the accretion-to-jet energy conversion efficiency in GW170817 alone, that the neutrino mechanism was responsible for the launching of the relativistic jet identified by \citet{Mooley2018a,Mooley2018b} and \citet{Ghirlanda2019}. 

\section{Discussion}

\begin{table}
 \caption{Summary of the accretion-to-jet energy conversion efficiencies discussed in the text. The first two entries represent the efficiency measured in this work. Other entries represent theoretical expectations for the \citet{Blandford1977} and $\nu\bar\nu$ annihilation mechanisms, either applied directly to GW170817 or based on simulations of generic BNS mergers. When applicable, we report both the median value and the 68\% confidence interval. }
 
 \centering
 \begin{tabular}{lcc}
  GW170817 measured & $\eta/10^{-3}$ & 68\% C.I.\\
  \hline
  Any $M_\mathrm{disc}$ & 1.2 & 0.2 -- 8.2\\
  $M_\mathrm{disc}>0.04\,\mathrm{M_\odot}$ & 0.6 & 0.1 -- 2.6 \\
  \hline\\
  Blandford-Znajek, fitting formulae$^a$ & ~ \\
  \hline
  \citet{McKinney2005} & 0.8 & 0.5 -- 1.7 \\
  \citet{Hawley2007} & 6.8 & 5.8 -- 7.0 \\
   MAD$^{b}$ & 380 & 375 -- 381 \\
   \hline\\
   
   Blandford-Znajek, BNS simulations\\
   \hline
   \citet[][BT]{Christie2019}  & $\sim$ 0.46 & ~ \\
   \citet[][BPW]{Christie2019} & $\sim$ 9.4  & ~ \\
   \citet[][BPS]{Christie2019} & $\sim$ 28   & ~ \\ 
   \citet[][]{Ruiz2019,Ruiz2020} & $\sim$ 2--3$^c$   & ~ \\ 
  \hline\\
  $\nu\bar\nu$ mechanism, GW170817 & ~ \\
  \hline\\ 
  \citet{Zalamea2011} & 0.40 & 0.19 -- 0.58 \\
  \hline
 \end{tabular}
 \flushleft\footnotesize{$^a$Based on GRMHD simulations of geometrically thick accretion discs around rotating BHs, seeded with poloidal
 magnetic fields (see the caveats in Sect.\ S\ref{sec:BZ_eff}). The fitting formulae are evaluated at the GW170817 remnant BH spin (see Appendix~\ref{sec:BH_mass_and_spin}).\\ $^b$Based on \citet{Tchekhovskoy2015}.\\$^c$These values represent the instantaneous efficiency $L_\mathrm{jet}/\dot M c^2$ at the end of the simulation.}
\label{tab:efficiencies}
 \end{table}

In Table~\ref{tab:efficiencies}, we summarise  the values and confidence ranges of our estimates of the accretion-to-jet energy conversion efficiency in GW170817, as well as the theoretically expected values for the \citet{Blandford1977} and $\nu\bar\nu$ jet-launching mechanisms. Our estimates are based on a comparison of the jet energy obtained by fitting the GRB 170817 afterglow and VLBI centroid motion by \citet{Ghirlanda2019} with the GW170817 disc mass obtained by applying the fitting formula by \citet{Barbieri2020} to the publicly available posterior samples from GW parameter estimation \citep{Abbott2019}. We list two estimates that were obtained either requiring the accretion disc mass to be at least $0.04\,\mathrm{M_\odot}$ \citep[][based on the large ejectum mass inferred from observations of the AT2017gfo kilonova, interpreted as being produced by disc winds]{Margalit2017,Radice2019} or imposing no constraints on the disc mass. These estimates point to a rather low efficiency of $\sim 10^{-3}$, which is not unexpected: Owing to the predominantly toroidal magnetic field configuration in the disc right after the merger \citep{Kiuchi2014} and to the relatively short duration of the accretion, the \citet{Blandford1977} process in the BNS post-merger is expected to be rather inefficient \citep{Christie2019,Ruiz2019,Ruiz2020}. Such a low efficiency is also compatible, in principle, with that of the neutrino-antineutrino annihilation mechanism, even though direct simulations of BNS mergers \citep{Ruffert1999,Just2016} seem to indicate that the baryon pollution in the polar region of the post-merger system, at the time when most of the $\nu\bar\nu$ energy is deposited, is too high for the jet to successfully propagate and break out.

Such a low accretion-to-jet energy conversion efficiency contrasts with the much higher values derived for flat-spectrum radio quasars and blazars by several authors \citep[e.g.][]{Rawlings1991,Ghisellini2014,Pjanka2017,Soares2020}. This could be explained by a nearly maximal BH spin in these systems \citep[e.g.][while in binary neutron star merger remnants, typically $a_\mathrm{BH}\sim 0.7$, see Appendix~\ref{sec:BH_mass_and_spin}]{Soares2020},  given the steep dependence of the efficiency on this parameter found in simulations of the BZ process \citep{McKinney2005,Hawley2007,Tchekhovskoy2010,McKinney2012}. Another difference that may play a role is the accretion rate: While GRB central engines accrete at extremely super-Eddington rates, supermassive BHs in quasars are thought to typically accrete at or near the Eddington limit \citep[e.g.][]{Maraschi2003}. Clearly, the large difference could instead be due to two different jet-launching mechanisms at play. Last, but not least, we note that the efficiency found in this work represents an average value over the entire duration of the accretion and that the instantaneous efficiency likely varies significantly over time (see Sect. \ref{sec:BZ_eff}), which makes the comparison with quasars more difficult.

Our efficiency estimate is in principle model-dependent as it relies on afterglow modelling as a means to measure the jet energy, as well as on a fitting formula based on a limited number of general-relativistic numerical simulations (which in turn suffer from limited resolution, uncertainty on the EoS at supra-nuclear densities, and the difficulty in accounting for the effects of neutrinos
and magnetic fields at the same time) as a means to estimate the disc mass. Nevertheless, as noted before, the disc mass cannot be much less than the $\sim 0.1\,\mathrm{M_\odot}$ most probable value derived using the \citet{Barbieri2020} fitting formula, given the constraints imposed by the large kilonova mass \citep{Margalit2017}. Moreover, we account to some extent for systematic uncertainties by introducing a relatively large dispersion in the values of the disc mass predicted by the fitting formula (Sect. \ref{sec:disc_mass}), and we properly account for statistical uncertainties -- including those arising from intrinsic afterglow model degeneracies (such as those pointed out by \citealt{Nakar2020}) -- in the jet energy
by taking the full posterior 
from \citet{Ghirlanda2019}. We note, moreover, that our adopted jet energy estimate is in agreement with essentially all others in the literature (see Fig.~\ref{fig:jet_energy_posterior}), some of which are based on significantly different models (e.g.~in terms of the adopted jet structure).  We are therefore confident that our conclusions are not heavily affected by systematics.

\section{Conclusions}
In this work we obtained, for the first time, an estimate of the accretion-to-jet energy conversion efficiency in a GRB by combining independent measurements of the jet energy (from multi-wavelength afterglow modelling) and of the disc mass (by applying a fitting formula based on a large suite of numerical simulations to the binary parameters inferred from gravitational wave parameter estimations). The resulting efficiency is rather low, $\eta\sim 10^{-3}$ (with large error bars), in agreement with expectations from both the \citet{Blandford1977} jet-launching mechanism, in the presence of a predominantly toroidal magnetic field configuration in the disc right after the merger, and the $\nu\bar\nu$ annihilation mechanism; the efficiency therefore does not allow us to distinguish between the two. Future applications of this method to a larger number of systems with well-measured gravitational wave parameters and well-sampled jet afterglows, together with improved disc mass predictions from numerical relativity simulations of binary neutron star mergers, will reduce the uncertainty in the efficiency and reveal its distribution
among different systems. Such a distribution could be the key to distinguishing between the two candidate jet-launching mechanisms due to the likely different dependencies on the disc and BH remnant properties: While the efficiency in both mechanisms increases with the BH spin, the dependence on the disc mass is likely different in the two scenarios. In the $\nu\bar\nu$ case, a higher disc mass favours a higher initial accretion rate, which extends the duration of the high neutrino luminosity phase (Sect. \ref{sec:nunu}). In the BZ case, on the other hand, the role of disc mass is less clear: Depending on the origin of the material that forms the disc -- either resulting from the tidal disruption of one or both the neutron stars, or from other processes that take place during and immediately after the merger -- the seed magnetic field configuration in the disc could be different, potentially affecting the efficiency \citep{Christie2019}. While a detailed investigation of such a dependence is beyond the scope of this work, we expect it to differ significantly from the $\nu\bar\nu$ case. This could therefore constitute the basis to distinguish the two mechanisms from each other through the application of the method described in this work to a large sample of events. Independently of this, polarimetric observations of the SGRB prompt emission and, potentially, of the reverse
shock in the early afterglow phase \citep{Lamb2019revshock} could reveal the degree of magnetisation in the jet, which can provide further insights into the jet-launching mechanism.

Once the accretion-to-jet energy conversion efficiency in SGRBs (and its possible dependence on the progenitor binary parameters) is established, it will be possible to infer the distribution of disc masses in jet-launching binary neutron star mergers by converting SGRB energies, 
even in the absence of a GW detection.
The disc mass distribution, in turn, can shed light 
on the distribution of properties of the progenitor binaries \citep[e.g.][]{Giacomazzo2013}. A similar approach can also be applied to BH-neutron star mergers if they are also found to produce GRBs. In that case, the information on the accretion-to-jet energy conversion efficiency could be used to constrain the EoS of matter at supra-nuclear densities \citep{Ascenzi2019}.

\begin{acknowledgements}
We thank Oliver Boersma for pointing out a typographic error in Equation 1 in the published version of this paper, which is now corrected (both in the present arXiv version and as a corrigendum on A\&A). We thank the anonymous referee for insightful comments that helped to improve the quality and presentation of this work. We thank A. Celotti, G. Ghisellini, G. Ghirlanda, G. Oganesyan, S. Ascenzi, C. Barbieri, R. Ciolfi, A. Perego, D. Lazzati and G. Lamb for useful discussions and comments. O.~S.\ acknowledges the INAF-Prin 2017 (1.05.01.88.06) and the Italian Ministry for University and Research grant ``FIGARO'' (1.05.06.13) for support.
\end{acknowledgements}

\footnotesize{
\bibliographystyle{aa}
\bibliography{references}
}

\appendix

\section{Black hole mass and spin}\label{sec:BH_mass_and_spin}
 
\begin{figure}
 \includegraphics[width=\columnwidth]{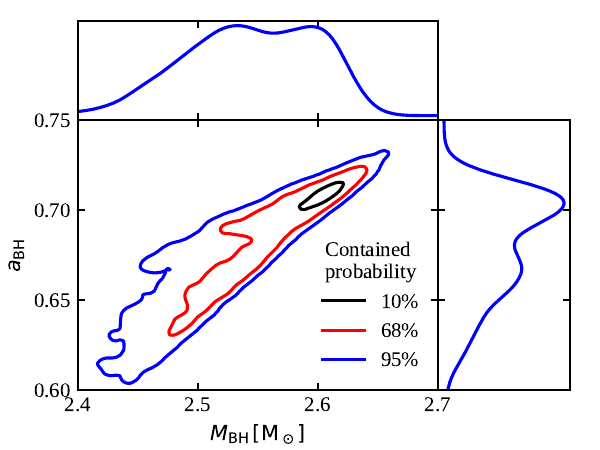}
 \caption{GW170817 remnant BH mass and spin posterior distributions, based on the \citet{Abbott2019} posterior samples from the GW parameter estimation and on the \citet{Coughlin2019} fitting formulae.}
 \label{fig:remnant_mass_spin}
\end{figure}

A crucial parameter that regulates the efficiency of both the neutrino and \citet{Blandford1977} mechanisms is the BH spin. In neutron star mergers, the spin of the final remnant is set by angular momentum conservation: The simplest estimate is obtained by equating the angular momentum of the remnant BH to the orbital angular momentum of the binary prior to merger, which suggests a typical value around $a_\mathrm{BH}\sim 0.7$. A more accurate estimate requires  that the angular momentum lost in gravitational waves and stored in bound and unbound matter that does not immediately accrete onto the BH be taken into account. Based on a series of general relativistic numerical simulations, \citet{Coughlin2019} provide a fitting formula  for the spin of the remnant BH in neutron star mergers, which depends on the component masses and on the dimensionless tidal deformability of the binary. The formula reads\footnote{The formula was reported incorrectly in \citet{Coughlin2019}, but a comparison with the BH remnant mass formula allows one to figure out the missing pieces.}
\begin{equation}
 a_\mathrm{BH} = \tanh\left[0.537 (4 \nu)^2 \left(\frac{M}{\mathrm{M_\odot}}-0.185\frac{\tilde\Lambda}{400}\right)-0.514\right]
 \label{eq:aBH_fit}
,\end{equation}
where $\nu = M_1 M_2/(M_1+M_2)^2$ and $M=M_1+M_2$. Based on the same suite of simulations, they also provide a fitting formula for the remnant BH mass, which reads
\begin{equation}
 M_\mathrm{BH} = 0.98 (4 \nu)^2 \left(\frac{M}{\mathrm{M_\odot}}-0.093\frac{\tilde\Lambda}{400}\right)
 \label{eq:MBH_fit}
.\end{equation}
Applying these formulae to the LVC posterior samples for the low spin prior, we obtain the posterior distributions shown in Fig.~\ref{fig:remnant_mass_spin}, which we then used as inputs to our \citet{Blandford1977} and $\nu\bar\nu$ annihilation efficiency calculations in Sects.\ \ref{sec:BZ_eff} and \ref{sec:nunu}.

\end{document}